\RequirePackage[l2tabu, orthodox]{nag}
\documentclass{stsci_report}

\usepackage{booktabs}
\usepackage{rotating}
\usepackage{natbib}
\usepackage{url}
\usepackage{graphicx}
\usepackage[font=footnotesize,labelfont=bf]{caption}
\usepackage{subcaption}
\usepackage{nameref}
\usepackage{array} 
\usepackage{color}
\usepackage{amsmath}

\usepackage[pdftex, pdfauthor={G. Anand}, pdftitle={Investigating ACS/WFC Amp-to-Amp Sensitivities}, pdfsubject={Recently, the ACS team applied an Ubercal framework to assess the photometric repeatability of stars observed across the WFC detector using 15 years of post-SM4 calibration data in the globular cluster 47 Tuc (Ryan et al., 2024). A surprising finding was an apparent 0.05 mag global difference in sensitivity between the WFC1 and WFC2 chips, which had not been seen in prior tests of sensitivity variations around the field-of-view. Given the many degenerate variables within the Ubercal framework such as CTE losses, time-dependent sensitivity, and flat-field corrections, we obtained new calibration data to perform a straightforward test of the reported 5$\%$ flux offset between detectors. We observed three white dwarf standards with three filters at four positions on the detector (each on a different amplifier), but with the same number of x and y pixel transfers to mitigate differential CTE-related effects. For the F606W and F814W filters, the agreements are good to 0.4$\%$ on average, and always 1$\%$ or better in individual cases. The consistency of these two filters over all three stars and the four dither positions provides very strong evidence against the large global sensitivity offset between WFC1 and WFC2 as seen in the Ubercal work. Larger variations seen in the bluer F435W filter are likely a result of a sensitivity of the flat field in that filter to underlying spectral type, warranting a future solution.}, pdfkeywords={ACS, CCD}]{hyperref}

\copyrighttext{Copyright \copyright \the\year\ The Association of Universities for Research in Astronomy, Inc. All Rights Reserved.}

\title{\textbf{Investigating ACS/WFC Amp-to-Amp Sensitivities}}
\presubtitle{\flushleft{\includegraphics[width=5cm]{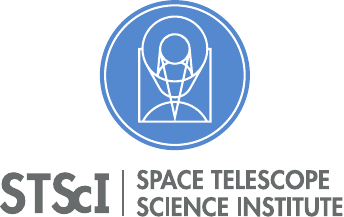}} \\ \hfill Instrument Science Report ACS 2026-03}
\author{G.S. Anand, N.A. Grogin}
\date{\today}

\begin{document}

\maketitle

\abstract{Recently, the ACS team applied an Ubercal framework to assess the photometric repeatability of stars observed across the WFC detector using 15 years of post-SM4 calibration data in the globular cluster 47 Tuc \citep{2024acs..rept....6R}. A surprising finding was an apparent 0.05~mag global difference in sensitivity between the WFC1 and WFC2 chips, which had not been seen in prior tests of sensitivity variations around the field-of-view. Given the many degenerate variables within the Ubercal framework such as CTE losses, time-dependent sensitivity, and flat-field corrections, we obtained new calibration data to perform a straightforward test of the reported $\sim$5$\%$ flux offset between detectors. 
We observed three white dwarf standards with three filters at four positions on the detector (each on a different amplifier), but with the same number of x and y pixel transfers to mitigate differential CTE-related effects. For the F606W and F814W filters, the agreements are good to 0.4$\%$ on average, and always 1$\%$ or better in individual cases. The consistency of these two filters over all three stars and the four dither positions provides very strong evidence against the large global sensitivity offset between WFC1 and WFC2 as seen in the Ubercal work. Larger variations seen in the bluer F435W filter are likely a result of a sensitivity of the flat field in that filter to underlying spectral type, warranting a future solution. }

\section*{Introduction} \label{s:intro}

The goal of the ACS/WFC absolute flux calibration program is to provide an accuracy at the level of 1$\%$ for external observers \citep{2016AJ....152...60B}. There are a variety of related instrumental calibrations that need to be incorporated within the flux calibration process to enable this 1$\%$ value, including but not limited to ongoing dark and bias corrections \citep{2025acs..rept....3G}, tracking of warm and hot pixels \citep{2020acs..rept....5M}, quantification of charge-transfer efficiency (CTE) losses \citep{2018acs..rept....4A, 2022acs..rept....6C}, changes in time-dependent sensitivity  \citep{2016AJ....152...60B}, and overall flat-fielding of the detector \citep{2002acs..rept....8M,2020acs..rept....1C}. 

The photometric repeatability of sources across the field-of-view (FOV) provides a test of these many calibrations. \cite{2012acs..rept....1B} used F775W observations from 2002 of the white dwarf standard G191B2B at seven positions around the field-of-view (FOV) and found a repeatability better than 0.5$\%$ at each position. A more comprehensive analysis was done by \cite{2015acs..rept....7B}, who tested the repeatability of both a blue, white dwarf standard (GD153) and a red, K-type standard (KF06T2) with the addition of 17 new detector positions. They find generally good ($\sim$1$\%$) agreement across the FOV in F814W. However, the agreement was notably worse in F435W for the blue standard (2.2$\%$) as opposed to the red standard (0.8$\%$), suggesting a possible dependence of spatial sensitivity on the underlying spectral type of the source. 

Putting aside the peculiarity with the F435W data (which we will revisit later in this text), there had been no compelling reason to suggest that the underlying sensitivity of the detector varies as a function of the two chips (WFC1 and WFC2) or four amplifiers (A$-$D). The two chips of the ACS/WFC detector were cut from the same silicon wafer, and their sensitivity profiles had shown to be continuous across the boundaries of the chip gap \citep{2025acsi.book...25S}. However, recently \cite{2024acs..rept....6R} undertook a new analysis within an Ubercal framework \citep{2008ApJ...674.1217P} that separates the issue of relative vs. absolute flux calibration, and attempts to provide a simultaneous fit for several parameters that impact photometric repeatability, including the low-frequency flat field, the magnitude of the time-dependent sensitivity, and any offsets between the four amplifiers. Their model was fit to 422 post-SM4 exposures of the 47~Tuc calibration field taken with the F606W filter. One key finding was a 0.05~mag, or $\sim$5$\%$ global flux sensitivity difference between the WFC1 (Amps A and B) and WFC2 (Amps C and D) chips. 

Given the null findings of prior studies, such an offset between the two chips comes as a surprise. Perhaps there could be a temporal component to this effect, as the last time this effect was studied with dedicated observations was over a decade ago. If this offset is genuine, then it must be accounted for to be able to provide the 1$\%$ overall ACS photometric calibration across the full FOV. Alternatively, given the simultaneous fitting of many (degenerate) variables performed within the Ubercal procedure, there is the possibility that this finding is an artifact of the process. In this report, we present a straightforward test of the possibility of chip-to-chip (or amp-to-amp) offsets within the WFC detector using observations of individual bright standard stars. 

\section*{Data and Processing} \label{data}

\begin{figure}[h!]
\centering
\includegraphics[width=0.80\linewidth]{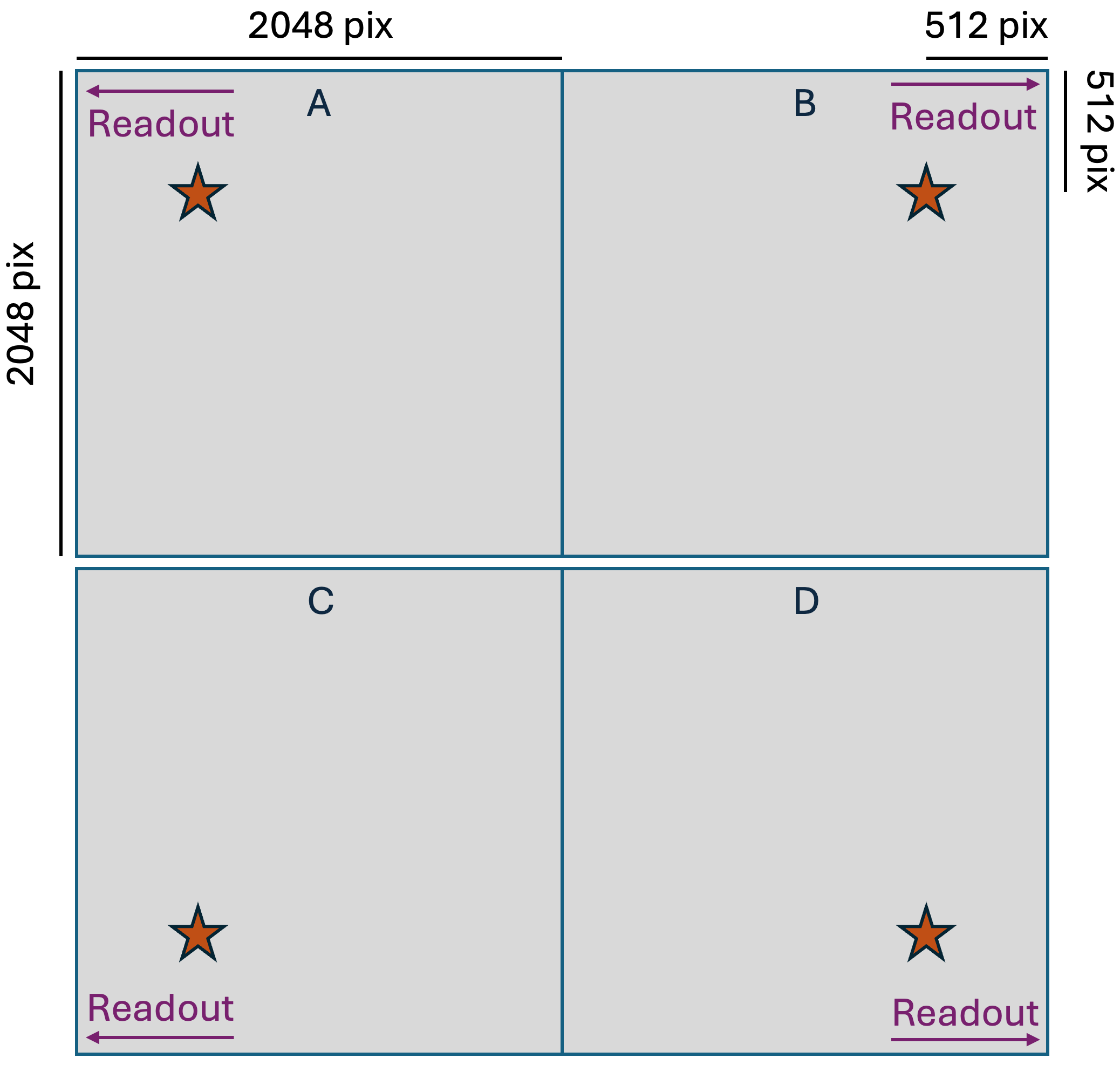}
\caption{Schematic of the observations used in this work. Three WD standards were observed on each of the four amplifiers, with their positions (stars) such that the number of x and y pixel transfers are identical, as shown by the 512 pixel scale bars at the top right.}
\label{fig:layout}
\end{figure}

To clarify the question of amp-to-amp sensitivity differences, we requested additional observations of three of the routine bright and isolated standard stars used in the yearly ACS photometric calibration program (CAL-17661; \citealt{2024hst..prop17661A}). In addition to the routine position on Amplifier B, we also observed these three white dwarfs (G191B2B, GD153, and GRW+70~5824) at the analogous position on each of the other three amplifiers, with the same number of pixel transfers (in both the x and y directions) to their respective readout locations. This setup allows us to negate any differential effects of either x (serial; \citealt{2024acs..rept....7R}) or y (parallel; \citealt{2018acs..rept....4A,2022acs..rept....6C}) CTE losses. These positions are indicated in Figure \ref{fig:layout}.


The stars were observed in each of the four positions using the F435W, F606W, and F814W filters. F606W was the filter used in the Ubercal work, and the two additional filters allows us to perform a test of any systematics as a function of wavelength. The exposure times are such that the overall photon noise ($\sim$0.3$\%$) for each source is minimal compared to the magnitude of the flux offset between chips ($\sim$5$\%$) found in the Ubercal work. Additionally, for each of the three stars, the time between individual visits was minimized (between 2~days and 6~weeks). Given that the magnitude of the time-dependent sensitivity losses are already much smaller (0.061$\%$/year; \citealt{2016AJ....152...60B}) than the size of the reported global chip-to-chip offset (5$\%$), any residual term from this effect should be entirely negligible. Overall, the straightforward observational design minimizes additional systematics and allows for a strong, clean test of any sensitivity differences between WFC1 and WFC2.

The data were taken as pairs of \textit{CR-SPLIT} exposures using the WFC-1K subarrays, and were manually processed through \texttt{acs$\_$destripe$\_$plus}. The resulting \textit{CRJ} files were used to perform aperture photometry with the standard 1$''$ = 20 pixel radius aperture, with a background annulus drawn at 6-8$''$ (past the infinite aperture radius). The pixel area maps for each observation were taken into account when computing the photometry. Given that the Amp B position shown in Figure \ref{fig:layout} is the standard for WFC flux calibration and thus the best calibrated position on the detector, we report the ratio of counts (or count rates, given the exposure times are identical for the four detector positions of each star+filter combination) relative to those determined on Amp B.

\section*{Results and Discussion} \label{s:results}

Results of the photometry are summarized in Tables \ref{tab:g191b2b}$-$\ref{tab:GRW+70}, with one table for each of the three standard stars. Results in each filter are shown in separate columns, and results for each amplifier (relative to amplifier B) are shown as individual rows. The final row provides the mean absolute error (MAE) over the three amplifier ratios, as a measure of average deviation from the ideal ratio of 1 (which would imply perfect agreement between amplifiers). Given that the individual photon noise is $\sim$0.3$\%$, the uncertainties in the individual ratios are $\sim$0.4$\%$, although this value does not take into account systematic uncertainties within the flat-fielding itself.

\begin{table}[]
\centering
\large
\begin{tabular}{|l|c|c|c|}
\hline
\textit{\textbf{}} & \textbf{F435W} & \textbf{F606W} & \textbf{F814W} \\ \hline
\textbf{A/B}               & 0.984          & 1.002          & 0.996          \\ \hline
\textbf{C/B}               & 0.999          & 0.999          & 1.003          \\ \hline
\textbf{D/B}               & 1.018          & 1.000          & 1.005          \\ \hline \hline \hline 
\textbf{\textit{MAE}}           & 1.2$\%$        & 0.1$\%$        & 0.4$\%$        \\ \hline
\end{tabular}
\caption{Summary of photometry for the white-dwarf standard G191B2B (DA0; B$-$V = $-$0.33). Rows show count rate ratios relative to the nominal Amp B position, whereas the columns show the results for three different filters. The final row provides the mean absolute error (calculated relative to the ``true" value of 1.000, which would imply perfect agreement), averaged separately over each filter.}
\label{tab:g191b2b}
\end{table}

\begin{table}[]
\centering
\large
\begin{tabular}{|l|c|c|c|}
\hline
\textit{\textbf{}} & \textbf{F435W} & \textbf{F606W} & \textbf{F814W} \\ \hline
\textbf{A/B}               & 0.987          & 1.006          & 0.997          \\ \hline
\textbf{C/B}               & 1.008          & 1.004          & 1.006          \\ \hline
\textbf{D/B}               & 1.031          & 1.004          & 1.009          \\ \hline \hline \hline 
\textbf{\textit{MAE}}           & 1.7$\%$        & 0.5$\%$        & 0.6$\%$        \\ \hline
\end{tabular}
\caption{Same as Table \ref{tab:g191b2b}, but for GD153 (DA1; B$-$V = $-$0.29)}
\label{tab:gd153}
\end{table}

\begin{table}[]
\centering
\large
\begin{tabular}{|l|c|c|c|}
\hline
\textit{\textbf{}} & \textbf{F435W} & \textbf{F606W} & \textbf{F814W} \\ \hline
\textbf{A/B}               & 0.985          & 1.002          & 0.997          \\ \hline
\textbf{C/B}               & 0.987          & 0.990          & 1.002          \\ \hline
\textbf{D/B}               & 1.007          & 0.993          & 1.007          \\ \hline \hline \hline 
\textbf{\textit{MAE}}           & 1.2$\%$        & 0.6$\%$        & 0.4$\%$        \\ \hline
\end{tabular}
\caption{Same as Table \ref{tab:g191b2b}, but for GRW+70~5824 (DA3; B$-$V = $-$0.09)}
\label{tab:GRW+70}
\end{table}

\begin{figure}[h!]
\includegraphics[width=\linewidth]{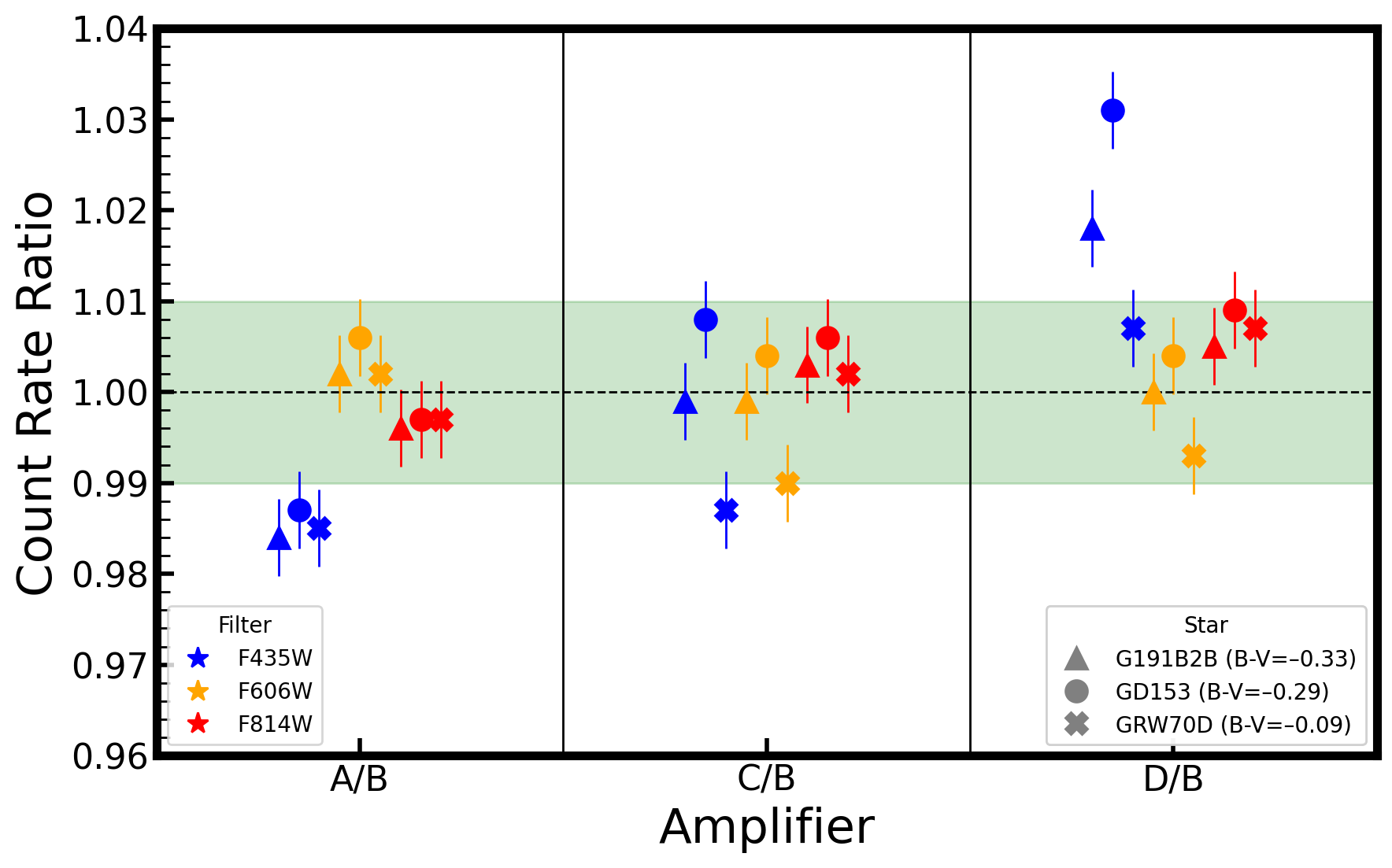}
\caption{Summary plot of the photometric results from this program. Data points are shown for each combination of individual stars (symbols), filters (colors), and amplifiers (columns). The green banding in the center shows a region of $\pm$1$\%$ as a visual aid.}
\label{fig:scatter}
\end{figure}

For the F606W and F814W filters, we see that the agreements are good to 0.4$\%$ on average, and always 1$\%$ or better in individual cases. The consistency of these two filters over all three stars and the four equivalent dither positions provides very strong evidence against the large global sensitivity difference of 5$\%$ between WFC1 and WFC2 as seen in the Ubercal work. However, in the bluer F435W filter, we find that the agreement is only good to 1.4$\%$ on average, and as poor as 3.1$\%$ in individual cases. The situation is summarized in Figure \ref{fig:scatter}, where the count rate ratios are shown for each combination of amplifier, standard star, and observing filter. The green banding highlights $\pm$1$\%$ as a simple visual aid. The number of discrepant data points in F435W is of immediate concern. This situation presents a substantial hurdle for obtaining 1$\%$ photometry around the FOV of the detector. Another curious observation is an apparent increase in scatter for the WFC2 amplifiers (C and D), which appears to decrease as a function of filter wavelength, in that the scatter is worst for the blue F435W filter.


We tested various adjustments to our analysis. Smaller extraction apertures provided somewhat noisier results, as expected due to breathing-induced focus-variations of the underlying point-spread functions \citep{2018acs..rept....8B}. We also tested different statistical criteria for the determination of the background levels, but these resulted in negligible differences, as the backgrounds are very low in these short exposures ($\leq$14~sec) of very bright standard stars.

\begin{figure}[h!]
\centering
\includegraphics[width=0.9\linewidth]{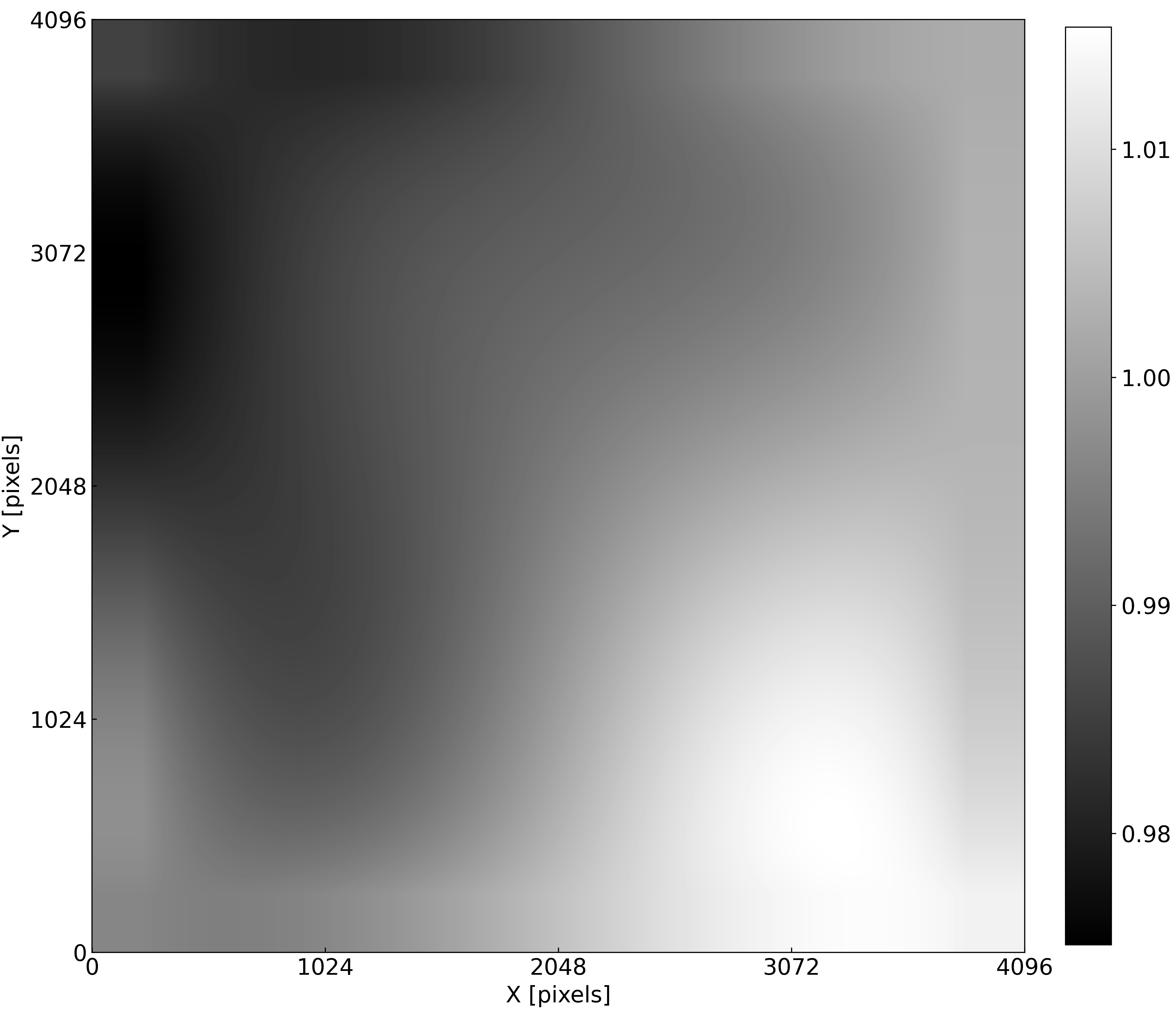}
\caption{Delta flat-field for blue sources observed in F435W (Bohlin et al. 2017). The values in this delta flat-field cover a range of 4$\%$, from 0.975 (black) to 1.015 (white).}
\label{fig:deltaff}
\end{figure}

So then, what could be the culprit for the discrepant results in F435W? We now return to the issue of a potential dependence of spatial sensitivity on the underlying spectral distributions of sources discussed in the Introduction. In other words, these prior works suggest that the F435W flat-field varies for sources of different colors. Most recently, \cite{2017acs..rept....9M} report that photometry of a red standard star shows a maximum deviation of 1.4$\%$ across the FOV, whereas the deviations for a blue standard are up to a substantial 2.5$\%$. They also point to an unpublished report (Bohlin et al. 2017) which developed a delta-flat correction for blue sources observed in the F435W filter. This delta flat-field is shown on the right-hand side of Figure 9 in \cite{2017acs..rept....9M}, and is reproduced here in Figure \ref{fig:deltaff}. The unpublished report also provided an estimate of the effects of serial CTE losses. While the report remains unpublished due to skepticism regarding the large magnitude of the proposed serial CTE correction, the delta-flat provides a potential avenue towards reducing the issues repeatedly seen for blue sources in F435W (while acknowledging that it is likely imperfect, given that the delta-flat analysis depended on estimated serial CTE losses)

\begin{figure}[h!]
\includegraphics[width=\linewidth]{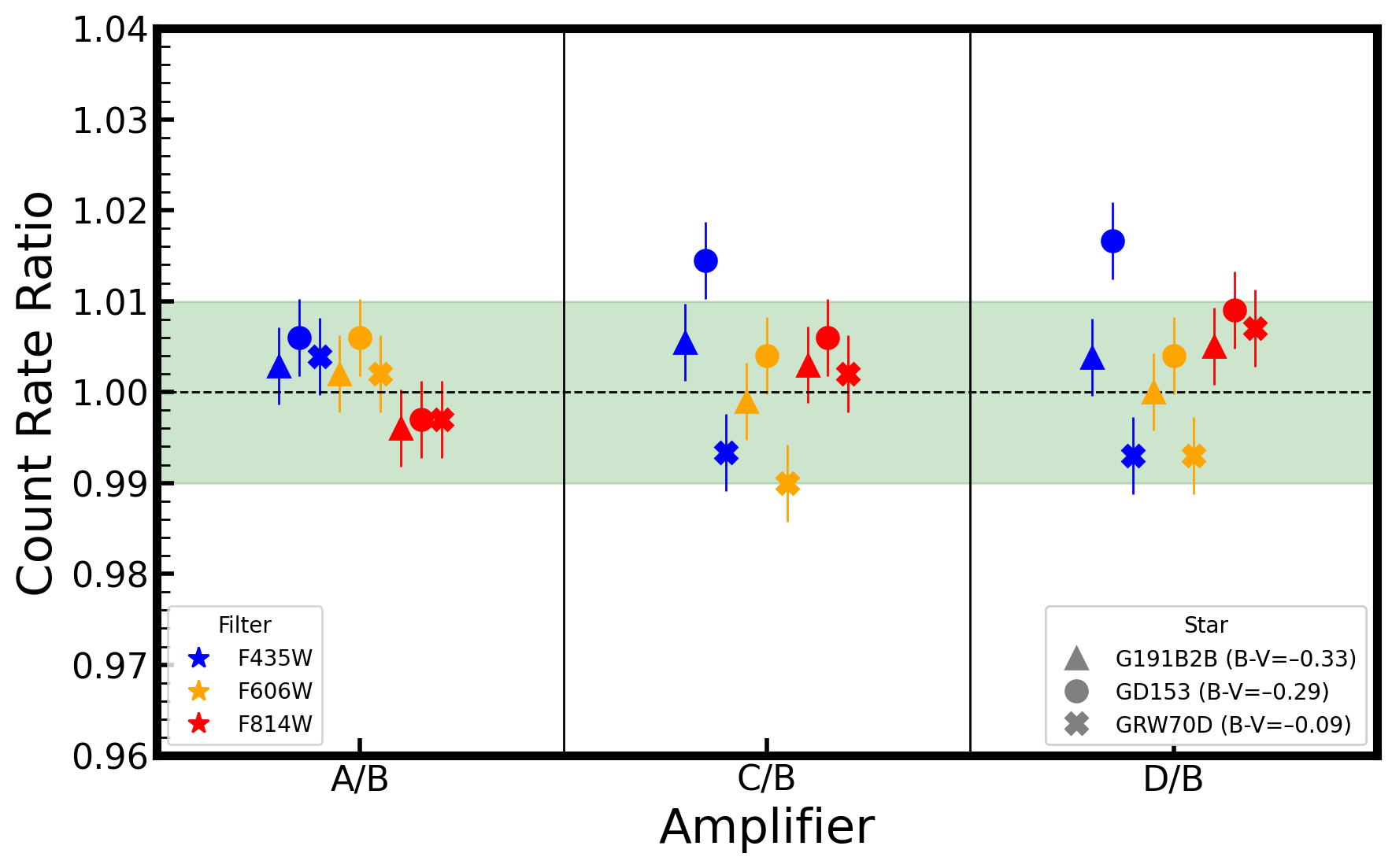}
\caption{Same as Figure \ref{fig:scatter}, but after application of the delta flat-field for blue sources observed in F435W (Figure \ref{fig:deltaff}), which improves the overall agreement across the FOV in that filter.}
\label{fig:scatterdeltaff}
\end{figure}

Applying the F435W delta-flat as a correction for the hot white dwarfs in our dataset, we arrive at the results in Figure \ref{fig:scatterdeltaff}, a simple variation of those shown in Figure \ref{fig:scatter}. The outcome can be seen visually as better agreement for each amplifier relative to Amp B, and more quantitatively by a reduction in the mean absolute errors. The reduction in this quantity for each star in F435W is from 1.2$\%$ to 0.4$\%$ (G191B2B), 1.7$\%$ to 1.2$\%$ (GD153), and 1.2$\%$ to 0.6$\%$ (GRW+70~5824). The indication given by this test further supports the idea of a sensitivity of the F435W flat-field to the underlying source color, and that the prior delta-flat correction rectifies a significant portion of the issue. Again, we caution that given the concerns surrounding the precise serial CTE corrections determined within that same work, the details of this unpublished delta-flat need to be reworked with our improved current understanding of the effects of serial CTE \citep{2024acs..rept....7R}. However, this methodology appears to provide a promising path for future investigations of photometric repeatability for blue sources observed with F435W.


\section*{Summary and Future Outlook} \label{s:summary}

Recent work by the ACS team using an Ubercal framework \citep{2024acs..rept....6R} uncovered a potential concern regarding a global sensitivity difference across the two chips (WFC1 vs. WFC2) at the level of 0.05~mag, or 5$\%$ in flux. This level of disagreement is substantially greater than the overall 1$\%$ flux calibration goal for the WFC. Prior tests of sensitivity differences around the WFC FOV returned generally null ($<$1$\%$) results. Given that these previous tests were performed over a decade ago, we obtained new observations to investigate this potential systematic concern. We performed a straightforward test using three bright WD standard stars observed in three filters (F435W, F606W, F814W). These standards were observed at one position on each amplifier, each with the same number of x and y pixel transfers from readout to mitigate any effects due to differential CTE losses.

Results from the F606W+F814W observations in the four dither positions for all three standard stars show typical agreement at the level of 0.4$\%$. This finding allows us to rule out the large (5$\%$) global chip-to-chip sensitivity difference seen in the Ubercal work. However, we find issues using the bluer F435W filter, seemingly confirming prior findings that spatial sensitivity varies in this filter depending on the underlying spectral type. The use of an unpublished delta-flat field for blue sources in F435W allows us to reduce the discrepancies by a factor of $\sim$2, although we caution that this work needs to be revised using our current knowledge of serial CTE. However, given the improved agreement, we suggest that further investigations along this route are needed to enable a high-accuracy flux calibration for sources of all colors observed with F435W filter. Another open question is whether this is an issue isolated to the F435W filter, or a larger issue present with the detector itself. Similar observations in the blue F475W filter would provide useful insight into resolving this matter. 

\section*{Acknowledgements}

We thank the HST Mission Office for their investment in the ACS team's Cycle 32 expanded calibration program. We thank the entire ACS team for useful discussions. We thank Jay Anderson, Roberto Avila, Ralph Bohlin, Christopher Clark, Jenna Ryon, and David Stark for their comments on an earlier version of this report. We also thank Jay Anderson for assistance with the observational design of this program.

The following packages were used in this analysis: \texttt{acstools} \citep{2020ascl.soft11024L}, \texttt{astropy} \citep{astropy:2022}, \texttt{jupyter} \citep{Kluyver:2016aa}, \texttt{matplotlib} \citep{Hunter:2007}, \texttt{numpy} \citep{harris2020array}, and \texttt{pandas} \citep{mckinney-proc-scipy-2010}.

\bibliography{main}
\bibliographystyle{apj}

\end{document}